# Faster Algorithms for Finding and Counting Subgraphs

Fedor V. Fomin[*]    Daniel Lokshtanov[*]    Venkatesh Raman[†]    B. V. Raghavendra Rao[†]
Saket Saurabh[*]

**Abstract**

Given an input graph $G$ and an integer $k$, the $k$-PATH problem asks whether there exists a path of length $k$ in $G$. The counting version of the problem, #$k$-PATH asks to find the number of paths of length $k$ in $G$. Recently, there has been a lot of work on finding and counting $k$-sized paths in an input graph. The current fastest (randomized) algorithm for $k$-PATH has been given by Williams and it runs in time $\mathcal{O}^*(2^k)$ [*IPL, 2009*]. The randomized algorithm for finding a $k$-path in the input graph was recently generalized by Koutis and Williams for testing whether there exists a subgraph in the input graph which is isomorphic to a given $k$-vertex tree [*ICALP, 2009*]. Björklund, Husfeldt, Kaski, and Koivisto [*ESA, 2009*] gave a deterministic algorithm for #$k$-PATH running in time and space $\mathcal{O}^*(\binom{n}{k/2})$ on an input graph with $n$ vertices and gave a polynomial space algorithm running in time $\mathcal{O}^*(3^{k/2}\binom{n}{k/2})$.

In this paper we study a natural generalization of both $k$-PATH and $k$-TREE problems, namely, the SUBGRAPH ISOMORPHISM problem. In the SUBGRAPH ISOMORPHISM problem we are given two graphs $F$ and $G$ on $k$ and $n$ vertices respectively as an input, and the question is whether there exists a subgraph of $G$ isomorphic to $F$. We show that if the treewidth of $F$ is at most $t$, then there is a randomized algorithm for the SUBGRAPH ISOMORPHISM problem running in time $\mathcal{O}^*(2^k n^{2t})$. To do so, we associate a new multivariate Homomorphism polynomial of degree at most $k$ with the SUBGRAPH ISOMORPHISM problem and construct an arithmetic circuit of size at most $n^{\mathcal{O}(t)}$ for this polynomial. Using this polynomial, we also give a deterministic algorithm to count the number of homomorphisms from $F$ to $G$ that takes $n^{\mathcal{O}(t)}$ time and uses polynomial space. For the counting version of the SUBGRAPH ISOMORPHISM problem, where the objective is to count the number of distinct subgraphs of $G$ that are isomorphic to $F$, we give a deterministic algorithm running in time and space $\mathcal{O}^*(\binom{n}{k/2}n^{2p})$ or $\binom{n}{k/2}n^{\mathcal{O}(t \log k)}$. We also give an algorithm running in time $\mathcal{O}^*(2^k \binom{n}{k/2}n^{5p})$ and taking space polynomial in $n$. Here $p$ and $t$ denote the pathwidth and the treewidth of $F$, respectively. Thus our work not only improves on known results on SUBGRAPH ISOMORPHISM but it also extends and generalize most of the known results on $k$-PATH and $k$-TREE.

## 1 Introduction

In this paper we consider the classical problem of finding and counting a fixed pattern graph $F$ on $k$ vertices in an $n$-vertex host graph $G$, when we restrict the treewidth of the pattern graph $F$ by $t$. More precisely the problems we consider are the SUBGRAPH ISOMORPHISM problem and the #SUBGRAPH ISOMORPHISM problem. In the SUBGRAPH ISOMORPHISM problem we are given two graphs $F$ and $G$ on $k$ and $n$ vertices respectively as an input, and the question is whether there exists a subgraph in $G$ which is isomorphic to $F$. In the #SUBGRAPH ISOMORPHISM problem the objective is to count the number of distinct subgraphs of $G$ that are isomorphic to $F$. Recently #SUBGRAPH ISOMORPHISM, in particular when $F$ has bounded treewidth, has found applications in the study of biomolecular networks. We refer to Alon et al. [2] and references there in for further details.

[*]Department of Informatics, University of Bergen, Norway. {fedor.fomin|daniello|saket}@ii.uib.no.
[†]The Institute of Mathematical Sciences, Chennai, India. {vraman|bvrr}@imsc.res.in.

In a seminal paper Alon et al. [4] introduced the method of COLOR-CODING for the SUBGRAPH ISOMORPHISM problem, when the treewidth of the pattern graph is bounded by $t$ and obtained randomized as well as deterministic algorithms running in time $2^{\mathcal{O}(k)} n^{\mathcal{O}(t)}$. This algorithm was derandomized using $k$-perfect hash families. In particular, Alon et al. [4] gave a randomized $\mathcal{O}^*(5.4^k)$[1] time algorithm and a deterministic $\mathcal{O}^*(c^k)$ time algorithm, where $c$ a large constant, for the $k$-PATH problem, a special case of SUBGRAPH ISOMORPHISM where $F$ is a path of length $k$. Using this algorithm for $k$-PATH, Alon et al [4] also resolved a conjecture of Papadimitriou and Yannakakis [22] that for $k = \mathcal{O}(\log n)$, the $k$-PATH problem can be solved in polynomial time. There has been a lot of efforts in parameterized algorithms to reduce the base of the exponent of both deterministic as well as the randomized algorithms for the $k$-PATH problem. In the first of such attempts, Chen et al. [11] and Kneis at al. [19] independently discovered the method of DIVIDE AND COLOR and gave a randomized algorithm for $k$-PATH running in time $\mathcal{O}^*(4^k)$. Chen et al. [11] also gave a deterministic algorithm running in time $\mathcal{O}^*(4^{k+o(k)})$ using an application of universal sets. While the best known deterministic algorithm for $k$-PATH problem still runs in time $\mathcal{O}^*(4^{k+o(k)})$, the base of the exponent of the randomized algorithm for the $k$-PATH problem has seen a drastic improvement. Koutis [20] introduced an algebraic approach based on group algebras for $k$-PATH and gave a novel randomized algorithm running in time $\mathcal{O}^*(2^{3k/2}) = \mathcal{O}^*(2.83^k)$. Williams [24] augmented the approach of Koutis [20] with more random choices and several other ideas and gave the current fastest algorithm for $k$-PATH running in time $\mathcal{O}^*(2^k)$. The best known algorithms for finding a HAMILTON PATH, case $k = n$ for the $k$-PATH problem, in an $n$-vertex graph run $\mathcal{O}^*(2^n)$ time and are quite old [6, 14, 15]. Any significant improvement in the run time dependence on $k$ given by Williams' algorithm would imply a faster HAMILTON PATH algorithm, which has been an open problem for over forty years.

While there has been a lot of work on the $k$-PATH problem, there has been almost no progress on other cases of the SUBGRAPH ISOMORPHISM problem until this year. Amini et al. [5] introduced an inclusion-exclusion based approach in the classical COLOR-CODING and using it gave a randomized $5.4^k n^{\mathcal{O}(t)}$ time algorithm and a deterministic $5.4^{k+o(k)} n^{\mathcal{O}(t)}$ time algorithm for the SUBGRAPH ISOMORPHISM problem, when $F$ has treewidth at most $t$. Koutis and Williams [21] generalized their algebraic approach for $k$-PATH to $k$-TREE, a special case of SUBGRAPH ISOMORPHISM problem where $F$ is a tree on $k$-vertices, and obtained a randomized algorithm running in time $\mathcal{O}^*(2^k)$ for $k$-TREE. Our first result fills this gap and generalizes the results of Koutis and Williams [21] and Williams [24] for $k$-TREE and $k$-PATH respectively, to the case when the pattern graph $F$ has treewidth at most $t$. More precisely, we give a randomized algorithm for the SUBGRAPH ISOMORPHISM problem running in time $\mathcal{O}^*(2^k(nt)^t)$, when the treewidth of $F$ is at most $t$. In general our approach follows the road map suggested by Koutis and Williams [21] and Williams [24], which is based on reducing the problem to checking a multilinear term in a specific polynomial of degree at most $k$. Our first non-trivial contribution is a new polynomial of degree at most $k$, namely the Homomorphism polynomial, relating graph homomorphisms and injective graph homomorphisms for testing whether a graph contains a subgraph which is isomorphic to a fixed graph $F$. We show that if the treewidth of the pattern graph $F$ is bounded by $t$ then we can make an arithmetic circuit of size $\mathcal{O}^*((nt)^t)$ for the Homomorphism polynomial which combined with a result of Williams [24] yields our first theorem. In fact, what we have is an arithmetic formula and using this we give a deterministic algorithm to count homomorphisms from $F$ to $G$ which runs in time $\mathcal{O}^*((nt)^{2t})$ and takes space polynomial in $n$, when the treewidth of $F$ is $t$. This is not only crucial for our polynomial space algorithm for counting subgraphs but also substantially improves the space requirement, from $k^{t+1} \log n$ to polynomial in $n$ and $k$, of the previous algorithm for counting graph homomorphisms of Diaz et al. [12].

In the second part of the paper we consider the problem of counting the number of pattern subgraphs, that is, the #SUBGRAPH ISOMORPHISM problem. The algorithm given in [6, 14, 15] for finding a hamiltonian path can in fact count the number of hamiltonian paths in the input graph in time $\mathcal{O}^*(2^n)$. Hence, a natural question is whether we can solve the #SUBGRAPH ISOMORPHISM problem in $\mathcal{O}^*(c^k)$ time, when the $k$-vertex graph $F$ is of bounded treewidth or whether we can even solve the #$k$-PATH problem in $\mathcal{O}^*(c^k)$ time? Flum and Grohe [13] showed that the #$k$-PATH problem is #W[1]-hard and hence it is very unlikely that the #$k$-PATH problem can be solved in time $f(k)n^{\mathcal{O}(1)}$ where $f$ is any arbitrary function of $k$. In another negative result, Alon and Gutner [3] have shown that one can not hope to solve #$k$-PATH better than $\mathcal{O}(n^{k/2})$ using the method of COLOR-CODING. They show this by proving that any family $\mathcal{F}$ of "balanced hash functions" from $\{1, \ldots, n\}$ to $\{1, \ldots, k\}$, must have size $\Omega(n^{k/2})$. On the positive side, very recently Vassilevska and Williams [23] studied various counting problems and among other results gave an algorithm for the #$k$-PATH problem running in time $\mathcal{O}^*(2^k(k/2)!\binom{n}{k/2})$ and space polynomial in $n$. Björklund et al. [8] introduced the method of "meet-in-the-middle" and gave an algorithm for the #$k$-PATH problem running in time and space $\mathcal{O}^*(\binom{n}{k/2})$. They also gave an algorithm for #$k$-PATH problem running in time $\mathcal{O}^*(3^{k/2}\binom{n}{k/2})$ and polynomial space, improving on the polynomial space algorithm given in [23]. We extend these results to the #SUBGRAPH ISOMORPHISM problem, when the pattern graph $F$ is of bounded treewidth or pathwidth. And here also graph homomorphisms come into play. By making use of graph homomorphisms we succeed to extend the applicability of the meet-in-the-middle method to much more general structures than paths. Combined with other tools—inclusion-exclusion, the DISJOINT SUM problem, separation property of graph of bounded treewidth or pathwidth and the trimmed variant of Yate's algorithm presented in [7]—we obtain the following results. Let $F$ be a $k$-vertex graph and $G$ be an $n$-vertex graph of pathwidth $p$ and treewidth $t$. Then #SUBGRAPH ISOMORPHISM is solvable in times $\mathcal{O}^*(\binom{n}{k/2}n^{2p})$ and $\binom{n}{k/2}n^{\mathcal{O}(t \log k)}$ and space $\mathcal{O}^*(\binom{n}{k/2})$. We also give an algorithm for #SUBGRAPH ISOMORPHISM that runs in time $\mathcal{O}^*(2^k\binom{n}{k/2}n^{3p}t^{2t})$ (respectively $2^k\binom{n}{k/2}n^{\mathcal{O}(t \log k)}$) and takes polynomial space. Thus our work not only improves on known results on SUBGRAPH ISOMORPHISM of Alon et al. [4] and Amini et al. [5] but it also extends and generalize most of the known results on $k$-PATH and $k$-TREE of Björklund et al. [8], Koutis and Williams [21] and Williams [24].

---
[1]Throughout this paper $\mathcal{O}^*()$ notation hides factors polynomial in the instance size $n$ and the parameter $k$.

The main theme of both algorithms, for finding and for counting a fixed pattern graph $F$, is to use graph homomorphisms as the main tool. Counting homomorphisms between graphs has found applications in variety of areas, including extremal graph theory, properties of graph products, partition functions in statistical physics and property testing of large graphs. We refer to the excellent survey of Borgs et al. [9] for more references on counting homomorphisms. In [5], for the first time, it was used to design exact and parameterized algorithms. One of the main advantages of using graph homomorphisms is that in spite of their expressive power, graph homomorphisms between many structures can be counted efficiently. Secondly, it allows us to generalize various algorithm for counting subgraphs with an ease. We also combine counting homomorphisms with the recent advancements on computing different transformations efficiently on subset lattice. Our deterministic polynomial space algorithm for counting graph homomorphisms uses arithmetic formula and it appears that this method could be useful in designing polynomial space variant of other exact algorithms.

## 2 Preliminaries

Let $G$ be a simple undirected graph without self loops and multiple edges. We denote the vertex set of $G$ by $V(G)$ and the set of edges by $E(G)$. For a subset $W \subseteq V(G)$, by $G[W]$ we mean the subgraph of $G$ induced by $W$. We refer to Appendix 6.1 for the standard definitions of treewidth, pathwidth and nice tree decomposition.

**Graph Homomorphisms:** Given two graphs $F$ and $G$, a graph *homomorphism* from $F$ to $G$ is a map $f$ from $V(F)$ to $V(G)$, that is $f : V(F) \to V(G)$, such that if $uv \in E(F)$, then $f(u)f(v) \in E(G)$. Furthermore, when the map $f$ is injective, $f$ is called an *injective homomorphism*. Given two graphs $F$ and $G$, the problem of SUBGRAPH ISOMORPHISM asks whether there exists an injective homomorphism from $F$ to $G$. By $\hom(F, G)$, $\inj(F, G)$ and $\sub(F, G)$ we denote the number of homomorphisms from $F$ to $G$, the number of injective homomorphisms from $F$ to $G$ and the number of distinct copies of $F$ in $G$, respectively. We denote by $\text{aut}(F, F)$ the number of automorphisms from $F$ to itself, that is bijective homomorphisms. The set $\text{HOM}(F, G)$ denotes the set of homomorphisms from $F$ to $G$.

**Functions on the Subset Lattice:** For two functions $f_1 : D_1 \to R_1$ and $f_2 : D_2 \to R_2$ such that for every $x \in D_1 \cap D_2$, $f_1(x) = f_2(x)$ we define the *gluing* operation $f_1 \oplus f_2$ to be a function from $D_1 \cup D_2$ to $R_1 \cup R_2$ such that $f_1 \oplus f_2(x) = f_1(x)$ if $x \in D_1$ and $f_1 \oplus f_2(x) = f_2(x)$ otherwise.

For a universe $U$ of size $n$, we consider functions from $2^U$ (the family of all subsets of $U$) to $\mathbb{Z}$. For such a function $f : 2^U \to \mathbb{Z}$, the zeta transform of $f$ is a function $f\zeta : 2^U \to \mathbb{Z}$ such that $f\zeta(S) = \sum_{X \subseteq S} f(X)$. Given $f$, computing $f\zeta$ using this equation in a naïve manner takes time $\mathcal{O}^*(3^n)$. However, one can do better, and compute the zeta transform in time $\mathcal{O}^*(2^n)$ using a classical algorithm of Yates [25]. In this paper we will use a "trimmed" variant of Yates's algorithm [7] that works well when the non-zero entries of $f$ all are located at the bottom of the subset lattice. In particular, it was shown in [7] that if $f(X)$ only can be non-zero when $|X| \leq k$ then $f\zeta$ can be computed from $f$ in time $\mathcal{O}^*(\sum_{i=1}^k \binom{n}{i})$. In our algorithm we will also use an efficient algorithm for the DISJOINT SUM problem, defined as follows. Input is two families $\mathcal{A}$ and $\mathcal{B}$ of subsets of $U$ and two weight functions $\alpha : \mathcal{A} \to \mathbb{Z}$ and $\beta : \mathcal{B} \to \mathbb{Z}$. The objective is to calculate

$$\mathcal{A} \boxtimes \mathcal{B} = \sum_{A \in \mathcal{A}} \sum_{B \in \mathcal{B}} \begin{cases} \alpha(A)\beta(B) & \text{if } A \cap B = \emptyset \\ 0 & \text{if } A \cap B \neq \emptyset \end{cases}$$

Following an algorithm of Kennes [16], Björkund et al. [8] gave an algorithm to compute $\mathcal{A} \boxtimes \mathcal{B}$ in time $\mathcal{O}(n(|\downarrow \mathcal{A}| + |\downarrow \mathcal{B}|))$, where $\downarrow \mathcal{A} = \{X : \exists A \in \mathcal{A}, X \subseteq A\}$ is the *down-closure* of $\mathcal{A}$.

**Arithmetic Circuits and Formula:** An arithmetic circuit (or a straight line program) $C$ over a specified ring $\mathbb{K}$ is a directed acyclic graph with nodes labeled from $\{+, \times\} \cup \{x_1, \ldots, x_n\} \cup \mathbb{K}$, where $X = \{x_1, \ldots, x_n\}$ are the input variables of $C$. Nodes with zero out-degree are called output nodes and those with labels from $X \cup \mathbb{K}$ are called input nodes. The *Size* of an arithmetic circuit is the number of gates in it. The *Depth* of $C$ is the length of the longest path between an output node and an input node. A *formula* is an arithmetic circuit where every node has out-degree bounded by 1, that is, the underlying undirected graph is a tree. The nodes in $C$ are sometimes referred to as *gates*. It is not hard to see that with every output gate $g$ of the circuit $C$ we can associate a polynomial $f \in \mathbb{K}[x_1, \ldots, x_n]$. For more details on arithmetic circuits see [10, 1].

A polynomial $f \in \mathbb{K}[x_1, \ldots, x_n]$ is said to have a multilinear term if there is a term of the form $c_S \prod_{i \in S} x_i$ with $c_S \neq 0$ and $\emptyset \neq S \subseteq \{1, \ldots, n\}$ in the standard monomial expansion of $f$.

## 3 Algorithm for Finding a Subgraph

In this section we give our first result and show that the SUBGRAPH ISOMORPHISM problem can be solved in time $\mathcal{O}^*(2^k(nt)^t)$ when the pattern graph $F$ has treewidth at most $t$. The main idea of our algorithm follows that of Koutis and Williams [21] and Williams [24] for the $k$-TREE problem and the $k$-PATH problem, respectively. However, we need additional ideas for our generalizations. Our second result of this section is a polynomial space algorithm to count graph homomorphisms between $F$ and $G$ when the treewidth of $F$ is at most $t$.

First, given two graphs $F$ and $G$, we will associate a polynomial $\mathcal{P}_G(X)$ where $X = \{x_v \mid v \in V(G)\}$ such that: (a) the degree of $\mathcal{P}_G(X)$ is $k$; (b) there is a one to one correspondence between the monomials of $\mathcal{P}_G$ and homomorphisms between $F$ and $G$; and (c) $\mathcal{P}_G$ contains a multilinear monomial of degree $k$ if and only if $G$ contains a subgraph isomorphic to $F$. The polynomial we associate with $F$ and $G$ to solve the SUBGRAPH ISOMORPHISM problem is given by the following.

$$\text{Homomorphism Polynomial} = \mathcal{P}_G(x_1, \ldots, x_n) = \sum_{\Phi \in \text{HOM}(F, G)} \prod_{u \in V(F)} x_{\Phi(u)}.$$

We first show that $\mathcal{P}_G$ is "efficiently" computable by an arithmetic circuit.

**Lemma 1.** *Let $F$ and $G$ be given two graphs with $|V(F)| = k$ and $|V(G)| = n$. Then the polynomial $\mathcal{P}_G(x_1, \ldots, x_n)$ is computable by an arithmetic circuit of size $\mathcal{O}^*((nt)^t)$ where $t$ is the tree-width of $F$.*

*Proof.* Let $F, G, k, n$ and $t$ be as given in the lemma. Let $D = (U, T, r)$ be a *nice* tree decomposition of $F$ rooted at $r$. We define a polynomial $f_G(T, \tau, U_\tau, S, \psi) \in \mathbb{Z}[X]$, where

- $\tau$ is a node in $T$;
- $U_\tau \subseteq V(F)$ is the vertex subset associated with $\tau$;
- $S$ be a multi-set (an element can repeat itself) of size at most $t+1$ with elements from the set $V(G)$;
- $\psi : F[U_\tau] \to G[S]$ is a multiplicity respecting homomorphism between the subgraphs induced by $U_\tau$ and $S$ respectively; and
- $X = \{x_v | v \in V(G)\}$ is the set of variables.

Let $V_\tau$ denote the union of vertices contained in the bags corresponding to the nodes of subtree of $T$ rooted at $\tau$. At an intuitive level $f_G(T, \tau, U_\tau, S, \psi)$ represents the polynomial which contains sum of monomials of the form $\prod_{u \in V_\tau \setminus U_\tau} x_{\phi(u)}$, where $\phi$ is a homomorphism between $F[V_\tau]$ and $G$ consistent with $\psi$, that is, $\phi$ is an extension of $\psi$ to $F[V_\tau]$. Formally, the polynomial $f_G$ can be defined inductively by going over the tree $T$ bottom up as follows.

**Case 1 (base case):** The node $\tau$ is a leaf node in $T$. Since $V_\tau = U_\tau$, there is only one homomorphism between $F[V_\tau]$ and $G$ that is an extension of $\psi$, hence $f_G(T, \tau, U_\tau, S, \psi) = 1$.

**Case 2:** The node $\tau$ is a *join node*. Let $\tau_1$ and $\tau_2$ be the two children of $\tau$ and $T_1$ and $T_2$ denote the sub-trees rooted at $\tau_1$ and $\tau_2$ respectively. Note that $U_\tau = U_{\tau_1} = U_{\tau_2}$ and $(V_{\tau_1} \cap V_{\tau_2}) \setminus U_\tau = \emptyset$. Hence, any extension of $\psi$ to a homomorphism between $F[V_{\tau_1}]$ and $G$ is independent of an extension of $\psi$ to a homomorphism between $F[V_{\tau_2}]$ and $G$. Thus we have,

$$f_G(T, \tau, U_\tau, S, \psi) = f_G(T_1, \tau_1, U_{\tau_1}, S, \psi) f_G(T_2, \tau_2, S, U_{\tau_2}, \psi). \tag{1}$$

**Case 3:** The node $\tau$ is an *introduce* node in $T$, let $\tau_1$ be the only child of $\tau$, and $\{u\} = U_\tau \setminus U_{\tau_1}$. Also, let $T_1$ denote the sub-tree of $T$ rooted at $\tau_1$. In this case any extension of $\psi$ to a homomorphism between $F[V_\tau]$ and $G$ is in fact an extension of $\psi|_{U_{\tau_1}}$ and thus we get

$$f_G(T, \tau, U_\tau, S, \psi) = f_G(T_1, \tau_1, U_{\tau_1}, S \setminus \{\psi(u)\}, \psi|_{U_{\tau_1}}). \tag{2}$$

**Case 4:** The node $\tau$ is a *forget* node in $T$, and $\tau_1$ is the only child of $\tau$ in $T$. Now, $U_{\tau_1}$ contains an extra vertex along with $U_\tau$. Thus any extension of $\psi$ to a homomorphism between $F[V_\tau]$ and $G$ is a direct sum of an extension of $\psi$ to include $u$ and that of $V_{\tau_1}$, where $\{u\} = U_{\tau_1} \setminus U_\tau$. Define, $Y \triangleq \{v \mid v \in V(G), \forall w \in U_\tau, wu \in E(F) \implies \psi(w)v \in E(G)\}$. For $v \in Y$, let $\psi_v : U_{\tau_1} \to S \cup \{v\}$ be such that $\psi_v|_{U_\tau} = \psi$ and $\psi_v(u) = v$. Then,

$$f_G(T, \tau, U_\tau, S, \psi) = \begin{cases} \sum_{v \in Y} \left( f_G(T_1, \tau_1, U_{\tau_1}, S \cup \{v\}, \psi_v) x_v \right) & \text{if } Y \neq \emptyset \\ 0 & \text{otherwise.} \end{cases} \tag{3}$$

Let $\text{HOM}(U_r, G)$ denote the set of all homomorphisms between the subgraph of $F$ induced by $U_r$ and $G$. In order to consider all homomorphisms between $F$ and $G$, we run through all homomorphisms $\psi$ between $F[U_r]$ and $G$, and then compute $f_G(T, r, U_r, Image(\psi), \psi)$ multiplied by the monomial corresponding to $\psi$. Now we define,

$$\mathcal{H}_G(T, r, U_r) = \sum_{\psi \in \text{HOM}(U_r, G)} f_G(T, r, U_r, S_\psi, \psi) \left( \prod_{u \in U_r, v = \psi(u)} x_v \right) \tag{4}$$

where, we consider the set $S_\psi = \text{Image}(\psi)$ as a multi set. Now we need to show that $\mathcal{H}_G$ is efficiently computable and $\mathcal{P}_G = \mathcal{H}_G$. We first show that $\mathcal{H}_G$ is computable by an arithmetic circuit of size $\mathcal{O}^*((nt)^t)$.

**Claim 1.** *$\mathcal{H}_G(T, r, U_r)$ is a polynomial of degree $k$ and is computable by an arithmetic circuit of size $\mathcal{O}^*((nt)^t)$. Here $r$ is the root of the tree $T$.*

*Proof.* In the above definition of $f_G$, the only place where the degree of the polynomial increases is at forget nodes of $T$. The number of forget nodes in $T$ is exactly $k - |U_r|$. Thus the degree of any $f_G$ is $k - |U_r|$ and hence the degree of $\mathcal{H}_G$ is $k$.

From the definitions in Equations (1-4) above, $\mathcal{H}_G(T, r, U_r)$ can be viewed as an arithmetic circuit $C$ with $X = \{x_v | v \in V(G)\}$ as variables and gates from the set $\{+, \times\}$. Any node of $C$ is labeled either by a variables from $U$ or a function of the form $f_G(T, \tau, U_\tau, S, \psi)$. The size of the circuit is bounded by the number of possible labelings of the form $f_G(T, \tau, U_\tau, S, \psi)$, where $T$ and $U_\tau$ are fixed. But this is bounded by $|V(T)| \cdot n^{t+1} \cdot (t+1)^{t+1} = (nt)^{t+\mathcal{O}(1)} = \mathcal{O}^*((nt)^t)$. □

Next we show that $\mathcal{H}_G$ defined above is precisely $\mathcal{P}_G$ and satisfies all the desired properties.

**Claim 2.** *Let $\phi : V(F) \to V(G)$. Then $\phi \in \text{HOM}(F,G)$ if and only if the monomial $\prod_{u \in V(F)} x_{\phi(u)}$ has a non-zero coefficient in $\mathcal{H}_G(T, r, U_r)$. In other words, we have that*

$$\mathcal{H}_G(T, r, U_r) = \mathcal{P}_G(x_1, \ldots, x_n) = \sum_{\phi \in \text{HOM}(F,G)} \prod_{u \in V(F)} x_{\phi(u)}.$$

*Proof.* We first give the forward direction of the proof. Let $\phi \in \text{HOM}(F, G)$ and $\psi = \phi|_{U_r}$. We show an expansion of $\mathcal{H}_G(T, r, U_r)$ which contains the monomial $\prod_{u \in V(F)} x_{\phi(u)}$. We first choose the term $f_G(T, r, U_r, S_\psi, \psi) \times \prod_{u \in U_r} x_{\psi(u)}$. We expand $f_G(T, r, U_r, S_\psi, \psi)$ further according to the tree structure of $T$. We describe this in a generic way. Consider the expansion of $f_G(T', \tau, U_\tau, S, \chi)$. If $\tau$ is a join node we recursively expand both the sub polynomials according to Equation (1). When $\tau$ is an introduce node we use Equation (2). In the case when $\tau$ is a forget node, we first note that $Y \neq \emptyset$ (this is the same $Y$ as defined in Case 4) and also that $\phi(u) \in Y$, where $u \in U_\tau \setminus U_{\tau_1}$. The last assertion follows from the definition of $Y$. Here, we choose the term which contains $x_{\phi(u)}$, note that there exists exactly one such term and proceed recursively.

Let $M$ denote the monomial obtained by the above mentioned expansion. For any node $v \in V(G)$, we have $\deg_M(x_v) = |\phi^{-1}(v)|$, where $\deg_M(x_v)$ denotes the degree of the variable $x_v$ in the monomial $M$. To see this, in the tree decomposition $D$, a node $u \in V(F)$ enters the tree through a unique forget node and this is exactly where the variable $x_{\phi(u)}$ is multiplied. Thus we have $M = \prod_{u \in V(F)} x_{\phi(u)}$. Note that this expansion is uniquely defined for a given $\phi$.

For the reverse direction, consider an expansion $\rho$ of $\mathcal{H}_G(T, r, U_r)$ into monomials and let $M = \prod x_v^{d_v}$ be a monomial of $\rho$, where $\sum d_v = k$. We build a $\phi \in \text{HOM}(F,G)$ using $\rho$ and the structure of $T$. Let $f_G(T, r, U_r, S_\psi, \psi)$ be the first term chosen using Equation (4). For every $u \in U_r$ let $\phi(u) = \psi(u)$. Inductively suppose that we are at a node $\tau$ and let $T'$ be the corresponding subtree of $T$. In the case of Equations (1) and (2) there is no need to do anything. In the case of Equation (3), where $\tau$ is a forget node, with $u \in U_{\tau_1} \setminus U_\tau$. If the expansion $\rho$ chooses the term $f_G(T_1, \tau_1, U_{\tau_1}, S \cup \{v\}, \psi_v) \times x_v$, then we set $\phi(u) = v$.

It remains to show that the map $\phi : V(F) \to V(G)$ as built above is indeed a homomorphism. We prove this by showing that for any edge $uu' \in E(F)$ we have that $\phi(u)\phi(u') \in E(G)$. If $uu'$ is an edge such that both $u, u' \in U_r$ then we are done, as by definition $\phi|_{U_r} \in \text{HOM}(U_r, G)$ and thus $\phi$ preserves all the edges between the vertices from $U_r$. So we assume that at least one of the end points of the edge $uu'$ is not in $U_r$. By the property of tree decomposition there is a $\tau' \in T$ such that $\{u, u'\} \in U_{\tau'}$. Now since at least one of the endpoints of $uu'$ is not in $U_r$, there is a node on the path between $r$ and $\tau'$ such that either $u$ or $u'$ is forgotten. Let $\tau''$ be the first node on the path starting from $\tau'$ to $r$ in the tree $T$ such that it does not contain both $u$ and $u'$. Without loss of generality let $u \notin U_{\tau''}$ and thus $\tau''$ is a forget node which forgets $u$. At any forget node, since the target node $v$ is from the set $Y$, we have that $\phi$ preserves the edge relationships among the vertices in $U_{\tau''}$ and $u$. Now from Equation (3), the property of $Y$ and the fact that $u' \in U_{\tau''}$ we have that $\phi(u)\phi(u') \in E(G)$. □

Now by setting $\mathcal{P}_G(X) = \mathcal{H}_G(T, r, U_r)$ the lemma follows which concludes the proof. □

We also need the following proposition proved by Williams [24], which tests if a polynomial of degree $k$ has a multilinear monomial with non-zero coefficient in time $\mathcal{O}(2^k s(n))$ where $s(n)$ is the size of the arithmetic circuit.

**Proposition 1** ([24]). *Let $P(x_1, \ldots, x_n)$ be a polynomial of degree at most $k$, represented by an arithmetic circuit of size $s(n)$ with $+$ gates (of unbounded fan-in), $\times$ gates (of fan-in two), and no scalar multiplications. There is a randomized algorithm that on every $P$ runs in $\mathcal{O}(2^k s(n) n^{\mathcal{O}(1)})$ time, outputs "yes" with high probability if there is a multilinear term in the sum-product expansion of $P$, and always outputs "no" if there is no multilinear term.*

Lemma 1 and Proposition 1 together yield our first theorem.

**Theorem 1.** *Let $F$ and $G$ be two graphs on $k$ and $n$ vertices respectively and $\mathbf{tw}(F) \leq t$. Then, there is a randomized algorithm for the SUBGRAPH ISOMORPHISM problem that runs in time $\mathcal{O}^*(2^k (nt)^t)$.*

**Counting Homomorphisms:** We note that the polynomial $\mathcal{P}_G$ can be used to count the number of homomorphisms from $F$ to $G$. Let $\hom(F, G)$ denote the number of homomorphisms from $F$ to $G$. Then we have, $\hom(F, G) = \mathcal{P}_G(1, \ldots, 1)$. First we make the following observations:

- $\mathcal{H}_G$ can in fact be computed by an arithmetic formula $\Phi$ of size $n^{\mathcal{O}(t)}$. In fact the straightforward circuit construction described above gives a formula.
- The depth of $\Phi$ is bounded by $1 + $ depth of $T$ and the depth of $T$ is bounded by $\mathcal{O}(k)$.

We need the following proposition.

**Proposition 2** (Folklore). [⋆] [2] *Given an arithmetic formula $\Phi$ of depth $d$ and size $s$, $\Phi$ can be evaluated at $(1, \ldots, 1)$ in time $\mathcal{O}(s)$ and $\mathcal{O}(d(d + \log s))$ bits of space.*

A naïve implementation of the above procedure would require $n^{\mathcal{O}(t)} + \mathcal{O}(d(d + \log s))$ bits of space, as we may need to store the whole formula $\Phi$ in the memory for an evaluation. Here, we give an implementation which reduces the space requirement to $\mathcal{O}(n(\log k + t \log n))$ bits. The idea is not to store the entire formula, instead to have a space efficient algorithm that given $\langle u, i, T \rangle$ as an input, outputs $i^{th}$ child of a node $u$ in $\Phi$.

**Lemma 2.** [⋆] *Given a nice tree decomposition of $F$, and a label $\langle T, \tau, U, S, \psi \rangle$ of a node $u$ in the formula $\Phi$ the following can be computed in $\mathcal{O}((n+k)(\log k + t \log n))$ bits of space and time $\mathcal{O}(Size(\Phi))$: (a) the number of children of $u$ in $\Phi$; and (b) the label of the $i^{th}$ child of $u$ in $\Phi$.*

We can implement the algorithm given in Proposition 2 by using the algorithm of Lemma 2 whenever we need to access an edge of the arithmetic formula $\Phi$. Note that even though every call to the algorithm of Lemma 2 requires $\mathcal{O}((n+k)(\log k + t \log n))$ bits of space, this can be reused among different calls. Thus the total space requirement is $\mathcal{O}(k(k + t \log n) + n(\log k + t \log n))$. Also, the overall running time of the evaluation procedure for $\Phi$ will be $\mathcal{O}^*((nt)^t \cdot Size(\Phi)) = \mathcal{O}^*((nt)^{2t})$. Hence we have the following:

---
[2] Proofs of results labeled with [⋆] have been moved to the appendix due to space restrictions.

**Theorem 2.** *Let $F$ and $G$ be two graphs on $k$ and $n$ vertices respectively and $\mathbf{tw}(F) \leq t$. Then, the number of homomorphisms from $F$ to $G$, $\hom(F, G)$, can be computed in time $\mathcal{O}^*((nt)^{2t})$ and $\mathcal{O}(k^2 + (k+n)t \log n)$ bits of space.*

This is a substantial improvement on space compared to the $\mathcal{O}(k^{t+1} \log n)$ bound of [12]. In fact, our algorithm takes polynomial number of bits, irrespective of the treewidth $t$. However the running time of our algorithm is $\mathcal{O}^*((nt)^{2t})$ compared to the $\mathcal{O}^*(k^t)$ time bound of [12].

## 4 Algorithms for Counting Subgraphs

In this section, we give algorithms for the #SUBGRAPH ISOMORPHISM problem, when $F$ has either bounded treewidth or pathwidth.

### 4.1 Counting Subgraphs with Meet in The Middle

When $|V(F)| = k$, the pathwidth of $F$ is $p$ and $|V(G)| = n$, then the running time of our algorithm for #SUBGRAPH ISOMORPHISM is $\mathcal{O}(\binom{n}{k/2} n^{2p+\mathcal{O}(1)})$. Roughly speaking, our algorithm decomposes $V(F)$ into three parts, the left part $L$, the right part $R$, and the separator $S$. Then the algorithm guesses the position of $S$ in $G$, and for each such position counts the number of ways to map $L$ and $R$ into $G$, such that the mappings can be glued together at $S$. Thus our result is a generalization of the meet in the middle algorithm for #$k$-PATH in an $n$-vertex graph by Björklund et al. [8]. However, our algorithm differs from that of Björklund et al. [8] conceptually in two important points. First, we count the number of injective homomorphisms from $F$ to $G$ instead of counting the number of subgraphs of $G$ that are isomorphic to $F$. To get the number of subgraphs of $G$ that are isomorphic to $F$ we simply divide the number of injective homomorphisms from $F$ to $G$ by the number of automorphisms of $F$. The second difference is that we give an algorithm that given a $k$-vertex graph $F$ of pathwidth $p$ and an $n$-vertex graph $G$ computes in time $\mathcal{O}^*(\binom{n}{k} n^p)$ the number of injective homomorphisms from $F$ to $G[S]$ for every $k$-vertex subset $S$ of $G$. In the #$k$-PATH algorithm of Björklund et al. [8], a simple dynamic programming algorithm to count $k$-paths in $G[S]$ for every $k$-vertex subset $S$, running in time $\mathcal{O}^*(\binom{n}{k})$ is presented, however this algorithm does not seem to generalize to more complicated pattern graphs $F$. Interestingly, our algorithm to compute the number of injective homomorphisms from $F$ to $G[S]$ for every $S$ is instead based on inclusion-exclusion and the trimmed variant of Yates's algorithm presented in [7]. In order to implement the meet-in-the-middle approach, we will use the following fact about graphs of bounded pathwidth.

**Proposition 3** (Folklore). [$\star$] *Let $F$ be a $k$-vertex graph of pathwidth $p$. Then there exists a partitioning of $V(F)$ into $V(F) = L \uplus S \uplus R$, such that $|S| \leq p$, $|L|, |R| \leq k/2$ and no edge of $F$ has one endpoint in $L$ and the other in $R$.*

Let $V(F) = L \uplus S \uplus R$ be a partitioning of $V(F)$ as given by Proposition 3, and let $L^+ = L \cup S$ and $R^+ = R \cup S$. For a map $g : S \to V(G)$ and a set $S'$ such that $S \subseteq S'$ and a set $Q$ we define $\hom_g(F[S'], G[Q])$ to be the number of injective homomorphisms from $F[S']$ to $G[Q]$ coinciding with $g$ on $S$. Similarly we let $\text{inj}_g(F[S'], Q)$ to be the number of homomorphisms from $F$ to $G[Q]$ coinciding with $g$ on $S$. If we guess how an injective homomorphism maps $F[S]$ to $G$ we get $\text{inj}(F, G) = \sum_g \text{inj}_g(F, G)$, where the sum is taken over all injective maps $g$ from $S$ to $V(G)$. For a given map $g$, we define the set of families $\mathcal{L}_g = \{Q \subseteq V(G) : |Q| = |L|\}$ and $\mathcal{R}_g = \{Q \subseteq V(G) : |Q| = |R|\}$. The weight of a set $Q \in \mathcal{L}_g$ is defined as $\alpha_g^L(Q) = \text{inj}_g(F[L^+], G[Q \cup g(S)])$ and the weight of a set $Q \in \mathcal{R}_g$ is set to $\alpha_g^R(Q) = \text{inj}_g(F[R^+], G[Q \cup g(S)])$.

For any $Q_1 \in \mathcal{L}_g$ and $Q_2 \in \mathcal{R}_g$ such that $Q_1 \cap Q_2 = \emptyset$, if we take an injective homomorphism $h_1$ from $F[L^+]$ to $G[Q_1 \cup g(S)]$ coinciding with $g$ on $S$ and another injective homomorphism $h_2$ from $F[R^+]$ to $G[Q_2 \cup g(S)]$ coinciding with $g$ on $S$ and glue them together, we obtain an injective homomorphism $h_1 \oplus h_2$ from $F$ to $G$. Furthermore two homomorphisms from $F$ to $G$ can only be equal if they coincide on all vertices of $F$. Thus, if $Q'_1 \in \mathcal{L}_g$, $Q'_2 \in \mathcal{R}_g$ and $h'_1$ and $h'_2$ are injective homomorphisms from $F[L^+]$ to $G[Q'_1 \cup g(S)]$ and from $F[R^+]$ to $G[Q'_2 \cup g(S)]$ respectively we have that $h_1 \oplus h_2 = h'_1 \oplus h'_2$ if and only if $h'_1 = h_1$ and $h'_2 = h_2$. Also, for any injective homomorphism $h$ from $F$ to $G$ that coincides with $g$ on $S$ we can decompose it into an injective homomorphism $h_1$ from $F[L^+]$ to $G[S \cup Q_1]$ and another injective homomorphism $h_2$ from $F[R^+]$ to $G[S \cup Q_2]$ such that $Q_1 \in \mathcal{L}_g$, $Q_2 \in \mathcal{R}_g$ and $Q_1 \cap Q_2 = \emptyset$. Then $\mathrm{inj}_g(F, G) = \mathcal{L}_g \boxtimes \mathcal{R}_g$ and hence

$$\mathrm{inj}(F, G) = \sum_g \mathcal{L}_g \boxtimes \mathcal{R}_g \tag{5}$$

**Proposition 4** ([8, 16]). *Given two families $A$ and $B$ together with weight functions $\alpha : \mathcal{A} \to \mathbb{N}$ and $\beta : \mathcal{B} \to \mathbb{N}$ we can compute the disjoint sum $\mathcal{A} \boxtimes \mathcal{B}$ in time $\mathcal{O}(n(|\downarrow \mathcal{A}| + |\downarrow \mathcal{B}|))$ where $n$ is the number of distinct elements covered by the members of $\mathcal{A}$ and $\mathcal{B}$. Here $\downarrow \mathcal{A} = \{X : \exists A \in \mathcal{A}, X \subseteq A\}$.*

We would like to use Proposition 4 together with Equation (5) in order to compute $\mathrm{inj}(F, G)$. Thus, given the mapping $g : S \to V(G)$ we need to compute $\mathcal{L}_g$, $\mathcal{R}_g$, $\alpha_g^L$ and $\alpha_g^R$. Listing $\mathcal{L}_g$ and $\mathcal{R}_g$ can be done easily in $\binom{n}{k/2} + \binom{n}{k/2}$ time, so it remains to compute efficiently $\alpha_g^L$ and $\alpha_g^R$.

**Lemma 3.** *Let $G$ be an $n$-vertex graph, $F$ be a $\ell$-vertex graph of treewidth $t$, $S \subseteq V(F)$ and $g$ be a function from $S$ to $V(G)$. There is an algorithm to compute $\mathrm{inj}_g(F, G[Q \cup g(S)])$ for all $\ell - |S|$ sized subsets $Q$ of $V(G) \setminus g(S)$ in time $\mathcal{O}^*((\sum_{j=1}^{\ell-|S|} \binom{n}{j}) \cdot n^p)$.*

*Proof.* We claim that the following inclusion-exclusion formula holds for $\mathrm{inj}_g(F, G[Q \cup g(S)])$.

$$\mathrm{inj}_g(F, G[Q \cup g(S)]) = \sum_{X \subseteq Q} (-1)^{|T|-|X|} \mathrm{hom}_g(F, G[X \cup g(S)]) \tag{6}$$

To prove the correctness of Equation (6), we first show that if there is an injective homomorphism $f$ from $F$ to $G[Q \cup g(S)]$ coinciding with $g$ on $S$ then its contribution to the sum is exactly one. Notice that since $|S| + |Q| = |V(F)|$, all injective homomorphisms that coincide with $g$ on $S$ only contribute when $X = Q$ and thus are counted exactly once in the right hand side. Since we are counting homomorphisms, in the right hand side sum we also count maps which are not injective. Next we show that if a homomorphism $h$ from $F$ to $G[S \cup Q]$, which coincides with $g$ on $S$, is not an injective homomorphism then its total contribution to the sum is zero, which will conclude the correctness proof of the equation. Observe that since $h$ is not an injective homomorphism it misses some vertices of $Q$. Thus $h(V(F)) \cap Q = W$ for some subset $W \subset Q$. We now observe that $h$ is counted only when we are counting homomorphisms from $F$ to $G[X \cup g(S)]$ such that $W \subseteq X$. The total contribution of $h$ in the sum, taking into account the signs, is

$$\sum_{i=|W|}^{|Q|} \binom{|Q|-|W|}{i-|Q|}(-1)^{|Q|-i} = \sum_{i=0}^{|Q|-|W|} \binom{|Q|-|W|}{i}(-1)^{|Q|-|W|-i} = (1-1)^{|Q|-|W|} = 0.$$

Thus, we have shown that if $h$ is not an injective homomorphism then its contribution to the sum is zero, and hence Equation (6) holds.

Observe that since $|Q| = \ell - |S|$, we can rewrite $(-1)^{|Q|-|X|}$ as $(-1)^{\ell-|S|-|X|}$. Define $\gamma(X) = (-1)^{\ell-|S|-|X|} \mathrm{hom}_g(F, G[X \cup g(S)])$, then we can rewrite Equation (6) as follows: $\mathrm{inj}_g(F, G[Q \cup g(S)]) = \gamma\zeta(Q)$. We start by pre-computing a table containing $\gamma(Q')$ for every $Q'$ with $|Q'| \leq \ell - |S|$. To do this we need to compute $\mathrm{hom}_g(F, G[Q' \cup g(S)])$ for all subsets $Q'$ of $V(G) \setminus g(S)$ of size at most $\ell - |S|$. There are at most $\sum_{j=1}^{\ell-|S|} \binom{n}{j}$ such subsets, and for each subset $Q'$ we can compute $\hom_g(F, G[Q' \cup g(S)])$, and hence also $\alpha(Q')$ using the dynamic programming algorithm of Diaz et al. [12] in time $\mathcal{O}^*(n^p)$. Now, to compute $\gamma\zeta(Q)$ for all $Q \subseteq V(G) \setminus g(S)$ of size $\ell - |S|$ we apply the algorithm for the trimmed zeta transform (ALGORITHM Z) from [7]. This algorithm runs in time $\mathcal{O}^*(\sum_{j=1}^{\ell-|S|} \binom{n}{j})$. Thus the total running time of the algorithm is then $\mathcal{O}^*((\sum_{j=1}^{\ell-|S|} \binom{n}{j}) \cdot n^p)$. This concludes the proof. □

We are now in position to prove the main theorem of this section.

**Theorem 3.** *Let $G$ be an $n$-vertex graph and $F$ be a $k$-vertex graph of pathwidth $p$. Then we can solve the #SUBGRAPH ISOMORPHISM problem in time $\mathcal{O}^*(\binom{n}{k/2} n^{2p})$ and space $\mathcal{O}^*(\binom{n}{k/2})$.*

*Proof.* We apply Proposition 4 together with Equation (5) in order to compute $\inj(F, G)$. In particular, for every mapping $g : S \to V(G)$ we list $\mathcal{L}_g$ and $\mathcal{R}_g$ and compute $\alpha_g^L$ and $\alpha_g^R$ using the algorithm from Lemma 3. Finally, to compute $\mathcal{L}_g \boxtimes \mathcal{R}_g$ we apply Proposition 4.

The sum in Equation (5) runs over $\binom{n}{p} p! \leq n^p$ different choices for $g$. For each $g$, listing $\mathcal{L}_g$ and computing $\alpha_g^L$, and listing $\mathcal{R}_g$ and computing $\alpha_g^R$, takes $\mathcal{O}^*(\binom{n}{k/2-|S|} n^p)$ and $\mathcal{O}^*(\binom{n}{k/2} n^p)$ time respectively. Finally, computing $\mathcal{L}_g \boxtimes \mathcal{R}_g$ takes time $\mathcal{O}^*(\binom{n}{k/2})$. Thus the total running time for the algorithm to compute $\inj(F, G)$ is $\mathcal{O}^*(\binom{n}{k/2} n^{2p})$.

To compute the number of occurrences of $F$ as a subgraph in $G$, we use the basic fact that the number of occurrences of $F$ in $G$ is $\inj(F, G)/\aut(F)$ [5]. Since $\aut(F) = \inj(F, F)$ we can compute $\aut(F)$ using the algorithm for computing $\inj(F, G)$ in time $\mathcal{O}^*(\binom{k}{k/2} n^{2p}) = \mathcal{O}^*(2^k n^{2p})$. This concludes the proof of the theorem. □

### 4.2 Polynomial Space Algorithm

In this section we give a polynomial space variant of our algorithm presented in the previous section. Our proof is similar to the one described by Björklund et al.[8] for the #$k$-PATH problem. We will also need the following proposition which gives a relationship between $\inj(F, G)$ and $\hom(F, G)$.

**Proposition 5** ([5])**.** *Let $F$ and $G$ be two graphs with $|V(G)| = |V(F)|$. Then*

$$\inj(F, G) = \sum_{W \subseteq V(G)} (-1)^{|W|} \hom(F, G[V(G) \setminus W]) = \sum_{W \subseteq V(G)} (-1)^{|V|-|W|} \hom(F, G[W]).$$

**Theorem 4.** *Let $G$ be an $n$-vertex graph and $F$ be a $k$-vertex graph of pathwidth $p$. Then we can solve the #SUBGRAPH ISOMORPHISM problem in time $\mathcal{O}^*(\binom{n}{k/2} 2^k n^{3p} t^{2t})$ and polynomial space.*

*Proof.* By Equation (5) we know that $\inj(F, G) = \sum_g \mathcal{L}_g \boxtimes \mathcal{R}_g$. We first show how to compute $\mathcal{L}_g \boxtimes \mathcal{R}_g$ for a fixed map $g : S \to V(G)$. For brevity, we use the Iverson Bracket notation: $[P] = 1$ if $P$ is true, and $[P] = 0$ if $P$ is false.

$$\begin{aligned}
\mathcal{L}_g \boxtimes \mathcal{R}_g &= \sum_{M \in \mathcal{L}_g} \sum_{N \in \mathcal{R}_g} [M \cap N = \emptyset] \alpha_g^L(M) \beta_g^R(N) \\
&= \sum_{M \in \mathcal{L}_g} \sum_{N \in \mathcal{R}_g} \sum_{\{X \subseteq V(G), |X| \leq k/2\}} (-1)^{|X|} [X \subseteq M \cap N] \alpha_g^L(M) \beta_g^R(N) \\
&= \sum_{\{X \subseteq V(G), |X| \leq k/2\}} (-1)^{|X|} \sum_{M \in \mathcal{L}_g} \sum_{N \in \mathcal{R}_g} [X \subseteq M]][X \subseteq N] \alpha_g^L(M) \beta_g^R(N) \\
&= \sum_{\{X \subseteq V(G), |X| \leq k/2\}} (-1)^{|X|} \Big( \sum_{M \in \mathcal{L}_g, M \supseteq X} \alpha_g^L(M) \Big) \Big( \sum_{N \in \mathcal{R}_g, N \supseteq X} \beta_g^R(N) \Big) \\
&= \sum_{i=1}^{k/2} \sum_{\{X \subseteq V(G), |X|=i\}} (-1)^i \Big( \sum_{M \in \mathcal{L}_g, M \supseteq X} \alpha_g^L(M) \Big) \Big( \sum_{N \in \mathcal{R}_g, N \supseteq X} \beta_g^R(N) \Big) \qquad (7)
\end{aligned}$$

For every $M \in \mathcal{L}_g$, by Equation (6), we know that the following inclusion-exclusion formula holds for $\alpha_g^L(M)$.

$$\alpha_g^L(M) = \mathrm{inj}_g(F[L^+], G[M \cup g(S)]) = \sum_{M' \subseteq M} (-1)^{|M|-|M'|} \mathrm{hom}_g(F[L^+], G[M' \cup g(S)])$$

We can compute $\mathrm{hom}_g(F[L^+], G[M' \cup g(S)])$ in $\mathcal{O}^*((nt)^{2p})$ time and polynomial space using a variant of Theorem 2. For details please see Appendix 6.5. Hence, using this we can compute $\alpha_g^L(M)$ in time $\mathcal{O}^*(2^{|M|}(nt)^{2p})$. Similarly we can compute $\alpha_g^R(N)$ in time $\mathcal{O}^*(2^{|N|}(nt)^{2p})$ for every $N \in \mathcal{R}_g$. Now using Equation (7) we can bound the running time to compute $\mathcal{L}_g \boxtimes \mathcal{R}_g$ as follows:

$$\sum_{i=1}^{k/2} \left( \binom{n}{i}\binom{n-i}{|L|-i} \mathcal{O}^*(2^{|L|}(nt)^{2p}) + \binom{n}{i}\binom{n-i}{|R|-i} \mathcal{O}^*(2^{|R|}(nt)^{2p}) \right)$$

$$\leq \sum_{i=1}^{k/2} \left( 2^{k/2}\binom{n}{|L|} \mathcal{O}^*(2^{|L|}(nt)^{2p}) + 2^{k/2}\binom{n}{|R|} \mathcal{O}^*(2^{|R|}(nt)^{2p}) \right)$$

$$\leq \sum_{i=1}^{k/2} \left( \binom{n}{k/2} \mathcal{O}^*(2^k(nt)^{2p}) + \binom{n}{k/2} \mathcal{O}^*(2^k(nt)^{2p}) \right) = k\binom{n}{k/2} \mathcal{O}^*(2^k(nt)^{2p}).$$

This implies that the time taken to compute $\mathrm{inj}(F, G) = \sum_g \mathcal{L}_g \boxtimes \mathcal{R}_g$ is $\mathcal{O}^*(2^k \binom{n}{k/2} n^{3p} t^{2t})$, as the total number of choices for $g$ is upper bounded by $\binom{n}{p} p! \leq n^p$. Finally, to compute the number of occurrences of $F$ in $G$, we use the basic fact that the number of occurrences of $F$ in $G$ is $\mathrm{inj}(F, G)/\mathrm{aut}(F)$ [5] as in the proof of Theorem 3. We can compute $\mathrm{aut}(F) = \mathrm{inj}(F, F)$, using the polynomial space algorithm given by Proposition 5 for computing $\mathrm{inj}(F, G)$ and Theorem 2, in time $\sum_{i=1}^{k} \binom{k}{i} \mathcal{O}^*((kp)^{2p}) = \mathcal{O}^*(2^k k^{4p})$ and space polynomial in $k$. This concludes the proof of the theorem. □

Theorems 3 and 4 can easily be generalized to handle the case when $F$ has treewidth at most $t$ by observing that if $\mathbf{tw}(F) \leq t$ then $\mathbf{pw}(F) \leq (t+1)\log(k-1)$ [18] and that Theorem 2 works for graphs of bounded treewidth.

## 5 Conclusion

In this paper we considered the SUBGRAPH ISOMORPHISM problem and the #SUBGRAPH ISOMORPHISM problem and gave the best known algorithms, in terms of time and space requirements, for these problems when the pattern graph $F$ is restricted to graphs of bounded treewidth or pathwidth. Counting graph homomorphisms served as a main tool for all our algorithms. We combined counting graph homomorphisms with various other recently developed tools in parameterized and exact algorithms like meet-in-middle, trimmed variant of Yates's algorithm, the DISJOINT SUM problem and algebraic circuits and formulas to obtain our algorithms. We conclude with an intriguing open problem about a special case of the SUBGRAPH ISOMORPHISM problem. Can we solve the SUBGRAPH ISOMORPHISM problem in time $\mathcal{O}^*(c^n)$, $c$ a fixed constant, when the maximum degree of $F$ is 3?

**Acknowledgements.** We thank Mikko Koivisto for pointing us to an error in the previous version of this manuscript and for useful discussions.

# 6 Appendix

## 6.1 Treewidth, Pathwidth and Nice Tree-Decomposition

A *tree decomposition* of a (undirected) graph $G$ is a pair $(U, T)$ where $T$ is a tree whose vertices we will call *nodes* and $U = (\{U_i \mid i \in V(T)\})$ is a collection of subsets of $V(G)$ such that

1. $\bigcup_{i \in V(T)} U_i = V(G)$,
2. for each edge $vw \in E(G)$, there is an $i \in V(T)$ such that $v, w \in U_i$, and
3. for each $v \in V(G)$ the set of nodes $\{i \mid v \in U_i\}$ forms a subtree of $T$.

The $U_i$'s are called bags. The *width* of a tree decomposition $(\{U_i \mid i \in V(T)\}, T)$ equals $\max_{i \in V(T)}\{|U_i| - 1\}$. The *treewidth* of a graph $G$ is the minimum width over all tree decompositions of $G$. We use notation $\mathbf{tw}(G)$ to denote the treewidth of a graph $G$. When in the definition of the treewidth, we restrict ourselves to path, we get the notion of *pathwidth* of a graph and denote it by $\mathbf{pw}(G)$. We also need a notion of *nice* tree decomposition for our algorithm. A *nice* tree decomposition of a graph $G$ is a tuple $(U, T, r)$, where $T$ is a tree rooted at $r$ and $(U, T)$ is a tree decomposition of $G$ with the following properties:

1. $T$ is a binary tree.
2. If a node $\tau \in T$ has two children, say $\tau_1$ and $\tau_2$, and $U_\tau = U_{\tau_1} = U_{\tau_2}$ then it is called *join node*.
3. If a node $\tau$ has one child $\tau_1$, $|U_\tau| = |U_{\tau_1}| + 1$ and $U_{\tau_1} \subseteq U_\tau$ then it is called *introduce node*.
4. If a node $\tau$ has one child $\tau_1$, $|U_{\tau_1}| = |U_\tau| + 1$ and $U_\tau \subseteq U_{\tau_1}$ then it is called *forget node*.
5. If a node $\tau$ is a leaf node of $T$ then it is called *base node*.

Given a tree-decomposition of width $t$, one can obtain a nice tree-decomposition of width $t$ in linear time.

## 6.2 Proof of Proposition 2

*Proof.* Starting with the root node of $\Phi$ we do a depth-first evaluation. At any instance we need to store the labels of nodes along a path under exploration which will require $\mathcal{O}(d \log s)$ space. We also need to store the partial evaluations along the path under exploration. Since output of a depth $d$ formula with $x_1 = 1, \ldots, x_n = 1$ as inputs can be stored using most $d$ bits, we would require $\mathcal{O}(d)$ bits per node in the path. Hence $\mathcal{O}(d^2)$ bits are required for the whole path. Hence the total space required is $\mathcal{O}(d(d + \log s))$. □

## 6.3 Proof of Lemma 2

*Proof.* Given an input gate $u$ of $\Phi$ with the label $\langle T, \tau, U, S, \psi \rangle$, we can describe the algorithm as follows:

1. Check if $U = U_\tau$ and $\psi$ is indeed a homomorphism between $F[U_\tau]$ and $G[S]$ else FAIL.
2. If $\tau$ is a leaf node then set $count = 0$ and return.
3. If $\tau$ is a join node then the node $\langle T, \tau, U, S, \psi \rangle$ has two children in $\Phi$ (see equation 1) hence set $count = 2$. If $i \leq 2$, then return the label of the $i^{th}$ child according the canonical ordering.
4. If $\tau$ is an introduce node and $\tau_1$ is the only child of $\tau$ in $T$, then set $count = 1$ and if $i = 1$ then return $\langle T, \tau_1, U_{\tau_1}, S \setminus \{\psi(u)\}, \psi|_{U_{\tau_1}} \rangle$ else FAIL. (See equation 2.)
5. If $\tau$ is a forget node then set $count = |Y|$. If $i \leq |Y|$ then return the $i^{th}$ tuple $\langle T, \tau_1, U_{\tau_1}, S \cup \{v\}, \psi_v \rangle$ in the canonical ordering of $\left\{ \langle T, \tau_1, U_{\tau_1}, S \cup \{v\}, \psi_v \rangle \right\}_{v \in Y}$. (See equation 3.)

*Time and space analysis:* The running time of the algorithm is clearly upper bounded by $Size(\Phi) = \mathcal{O}^*((nt)^t)$. For space, note that we can store $T$ using $\mathcal{O}(kt \log n)$ bits as there are $\mathcal{O}(k)$ nodes in $T$, each containing a bag of size at most $t+1$. Storing a label of the form $\langle T, \tau, U, S, \psi \rangle$ requires $\mathcal{O}(\log k + t \log n + t \log n + 2t \log n) = \mathcal{O}(\log k + t \log n)$ many bits of space. Steps 1-4 of the algorithm can be implemented within this space bound. However, step 5 requires a storing of the set $R = \left\{ \langle T, \tau_1, U_{\tau_1}, S \cup \{v\}, \psi_v \rangle \right\}_{v \in Y}$ and then choosing the $i^{th}$ element in their lexicographic ordering. However as $|R| = |Y| \leq |V(G)| = n$, we can store all these labels in $\mathcal{O}(n(\log k + t \log n))$ bits of space. So, the overall space requirement is $\mathcal{O}((n+k)n(\log k + t \log n))$ bits. $\square$

## 6.4 Proof of Proposition 3

*Proof.* The vertices of a graph $F$ of pathwidth $p$ can be ordered as $v_1 \ldots v_k$ such that for any $i \leq k$ there is a subset $S_i \subseteq \{v_1 \ldots v_i\}$ with $|S_i| \leq p$, such that there are no edges of $F$ with one endpoint in $\{v_1 \ldots v_i\} \setminus S_i$ and the other in $\{v_{i+1}, \ldots v_k\}$. Such an ordering is obtained, for example, in [17]. Choose $L' = \{v_1 \ldots v_{\lceil k/2 \rceil}\}$, $S = S_{\lceil k/2 \rceil}$, $L = L' \setminus S$ and $R = \{v_{\lceil k/2 \rceil + 1} \ldots v_k\}$. Then $L$, $S$ and $R$ have the claimed properties. $\square$

## 6.5 Variant of Theorem 2

For the proof of Theorem 4, we need a variant of Theorem 2. In Theorem 4, we need $\hom_g(F, G[X \cup S])$ instead of $\hom(F, G[X \cup S])$. This can be achieved if we can compute the polynomial,

$$\mathcal{P}_G^g(x_1, \ldots, x_n) = \sum_{\Phi \in \text{HOM}_g(F,G)} \prod_{u \in V(F)} x_{\Phi(u)},$$

by an arithmetic formula of size $n^{\mathcal{O}(p)}$, where $\text{HOM}_g(F, G)$ is the set of all homomorphisms from $F$ to $G$ that coincide with the function $g : S \to V(G)$, for $S \subseteq V(F)$. We obtain the desired circuit by modifying the constructions given in Lemma 1 as follows:

- the sum in Equation (4) runs over $\text{HOM}_g(F, G[X \cup S])$; and
- in Equation (3) we ensure that the sum is restricted to $\psi_v$'s that agree with $g$ on $S \subseteq V(F)$.

With an argument similar to that of Lemma 1 and using the fact that $\mathbf{tw}(F) \leq \mathbf{pw}(F)$ we conclude as follows.

**Lemma 4.** *Let $F$ and $G$ be given two graphs with $|V(F)| = k$ and $|V(G)| = n$. Let $X, S \subseteq V(G)$ and $g : V(F) \to S$ be any function. Then the polynomial $\mathcal{P}_G^g(x_1, \ldots, x_n)$ is computable by an arithmetic formula of size $\mathcal{O}^*((np)^{2p})$ where $p$ is the pathwidth of $F$.*

Now, by applying Proposition 2 and Lemma 2 we get a polynomial space algorithm to compute $\hom_g(F, G[X \cup S])$.